\documentclass[doublespacing]{elsart}
\usepackage{graphicx}

\def\textbf#1{{\bf #1}}
\def\textit#1{{\it #1}}

\begin{document}

\begin{frontmatter}

\title{Scaling behavior in land markets}
\author[Tokyo]{Taisei Kaizoji}
  \ead{kaizoji@icu.ac.jp}
\address[Tokyo]{%
Division of Social Sciences, International Christian University, Osawa, Mitaka, Tokyo 181-8585, Japan}

\begin{abstract}
We present power law statistics on land markets, which no other studies have so far done. We analyze a database of {\it the assessed value of land} that is officially monitored and is made public by the Ministry of Land, Infrastructure and Transport Government of Japan. This is the largest database of Japan's land prices consisting of approximately 30,000 points for each year of a 6-year period (1995-2000)[25]. By analyzing the data of the posted land price we found the power law distributions of the land prices and of the relative prices of the land. The data fits to a very good degree of approximation the power law distributions. We also found that the price fluctuations are amplified with the level of the price. These results hold for the data for each of the 6 annual intervals. Our empirical findings give the conditions that any empirically accurate theories of land market have to satisfy. 
\end{abstract}


\end{frontmatter}
\section{Introduction}
Power law scaling consists of universal properties that characterize collective phenomena that emerge from complex systems composed of many interacting units. Power law scaling has been observed not only in physical systems, but also in economic and financial systems [1-22]. The discovery of this scaling behavior in economic and financial systems has shed new light on economics, and, in recent years, has led to the establishment of a new scientific field bridging economics and physics. 
Market economy land is one of the most important assets in capitalism. The movement of land prices has a strong influence on the economic behavior of individuals and firms [23,24]. Over the past few decades, a considerable number of studies have been conducted on the scaling behavior in economic systems. Nevertheless, no studies have ever tried to study the scaling behavior in a land market. Here, we introduce power law statistics in the distributions of land prices and of the relative price, as a new example of power law scaling in economic systems. 

\section{Empirical Results}
\subsection{The data}
In most other countries, land price is a residual of property price, calculated by the income approach: net income divided by yield, minus building cost. However, in Japan, property price, including that for income producing properties, usually comprises land price plus building price, which are calculated separately. Land prices are normally determined by market comparisons utilizing land price indices prepared by the Government; such as the assessed value of land by Ministry of Land, Infrastructure and Transport, whilst building prices are normally determined in the market through the cost approach. The assessed value of land is a price of a centare of a place as of January 1, evaluated by two real estate valuers [25]. The investigation is undertaken throughout the land once a year. We analyze a database of the assessed value of land covering the 6-year period between 1995-2000. The data for each of the 6 annual intervals contains the land prices in approximately 30,000 points. 

\subsection{The distribution of the land price}
We first study the probability distribution of land price. In Fig.1, we plot the cumulative probability distribution of land price (S) of 30,600 points in the year 1998, which is the log-log plot of the cumulative probability as a function of land price. The ordinate shows the cumulative probability in a log scale, that is, the probability of finding land of a price equal to, or higher than, x. It is apparent that this plot tends towards becoming a linear function in the high price range. We find that the tail of the cumulative probability distribution of the land price is described by a power law distribution, 
\begin{equation}
    P(S > x) \propto x^{-\alpha}.  
\end{equation}
The solid line in Fig. 1 represents this function. In the range of land prices higher than $ 2 \times 10^5 $ yen, the probability distribution follows a power low with exponent $ \alpha = 1.7 \pm 0.02 $ $ (R^2 = 0.99) $, as determined by ordinary least squares (OLS) regression in log-log coordinates. Here, $ R^2 $ denotes the coefficient of determination. The distribution fits the power law (1) very accurately in the high price range but it gradually deviates from (1) as the price becomes lower. As a further test of the strength of this result, I repeated the same analyses for the data of land prices for each of the periods between 1995 - 2000. The power law exponents were in the range of $ (1.53, 1.76) $. The results are represented in Table 1. Overall, the power law distribution of land prices is very strong. \par 

\begin{table}[hbtp]
\begin{center}
\begin{tabular}{c|cccc}
\hline
{\bf Year} & {\bf Data points} & {\bf Estimated $ \alpha $} & {\bf S.E.} & {\bf R2} \\ \hline
1995 & 30000 & 1.53 & 0.07 & 0.98 \\ 
1996 & 29590 & 1.6  & 0.06 & 0.98 \\
1997 & 29401 & 1.72 & 0.05 & 0.99 \\

1998 & 30600 & 1.76 & 0.04 & 0.99 \\
1999 & 30198 & 1.76 & 0.03 & 0.99 \\
2000 & 30287 & 1.75 & 0.02 & 0.99 \\ \hline
\end{tabular} 
\bigskip
\caption{{\small Power law exponents $ \alpha $ for the cumulative probability 
distribution of Japan's land prices. The results are obtained by using OLS regression on data. S.E. denotes standard errors, and R2 the coefficient of determination.}} 
\bigskip
\bigskip
\end{center}
\end{table}

\subsection{The distribution of the relative land price} 
We next analyze the probability distribution of the relative price of land. The relative price is, by definition, $ r = S(t+1)/S(t) $ where $ S(t) $ and $ S(t+1) $ are the prices of a given point in the years $t$ and $t+1$, respectively. Fig.2 shows the probability distribution of the relative price ($r$) from 1997 to 1998, which is plotted on the log-log coordinates. Note that the horizontal axis denotes the relative price change that is defined as $ (\ln r) $. As seen from the figure, the probability distribution of relative prices is consistent with a scaling power law asymptotic behavior, 

\begin{equation}
   P(r) \propto |r|^{- \beta}
\end{equation}
Using ordinary least squares regression in Fig. 2, we obtain estimates for the asymptotic slope 
\begin{equation}
\beta = \left\{ 
\begin{array}{rl} 
 26.11 \pm 0.47 & \quad\mbox{for $ r < r^* (R^2 = 0.99)$}, \\
 112.37 \pm 6.02 & \quad\mbox{for $ \geq r^* (R^2 = 0.98)$}
\end{array} \right.
\end{equation}
where $ r^* = 1.005 $. As a further test of the strength of these results, we repeated these analyses for 6-year periods between 1995 and 2000. Similar results were obtained for each year back through 1995 (Table 2). \par

\begin{table}[hbtp]
\begin{center}
\begin{tabular}{c|ccc|ccc}
\hline
{\bf Year}& {\bf lower tail $ \beta $}& {\bf S.E.} & {\bf R2} & {\bf upper tail $ \beta $}& {\bf S.E.} & {\bf R2} \\ \hline
1995 & 12.68 & 0.85 & 0.88 & 91.22 & 5.52 & 0.98 \\ 
1996 & 14.45 & 0.62 & 0.95 & 90.92 & 3.87 & 0.99 \\
1997 & 18.34 & 0.76 & 0.95 & 89.43 & 4.17 & 0.99 \\
1998 & 26.11 & 0.47 & 0.99 & 112.37 & 6.02 & 0.98 \\
1999 & 21.8 & 0.57 & 0.98 & 123.87 & 8.26 & 0.99 \\
2000 & 22.52 & 0.61 & 0.98 & 107.71 & 12.12 & 0.96 \\ \hline
\end{tabular}
\bigskip
\caption[Table2]{Power law exponents $ \beta $ for the probability distributions of the relative prices of land in Japan. The results are obtained by using OLS regression on data. S.E. denotes standard errors, and R2 the coefficient of determination.} 
\bigskip
\bigskip
\end{center}
\end{table}

Fig. 2, which indicates the probability distribution of relative prices from 1997 to 1998, contains a long tail to the left, so the negative observations extend over a wide range, but the positive do not. The average of the Japanese land prices continued to rise after World War II. Particularly in the period of the bubble economy in Japan (1987-1991), the land prices had extraordinarily risen. However, the average price peaked in 1992, and started to decline continuously thereafter. The land prices have fallen during the period of the collapse of the bubble economy. That is why the distribution function of relative prices in the years 1997 and 1998 is skewed to the left. Furthermore, this suggests that the distribution function of the relative prices was skewed to the right during the bubble economy period, in which the land prices rose rapidly. A more credible hypothesis is that the slant of the probability distribution of the relative prices was a response to the changes in the economic environment and over time. \par 
Here, we analyze the distributions of the relative price in more detail, in order to examine the effect of price. To this aim, we partition the points into three price-ranges, {\it low}, {\it medium}, and {\it high}, according to the logarithm of price $ s = \ln S(t) $. Fig. 3a indicates the conditional probability density functions on a log-log scale. The figure and Table 3 show clearly that points that belong to the high price-range have a higher standard deviation than points that belong to the low price range. Thus the conditional probability density functions, $ P(r | s) $, have different standard deviations, for the log price. 

\begin{table}[h]
\begin{center}
\begin{tabular}{c|ccc|ccc|c}
\hline
{\bf Price range} & {\bf lower tail} $ \lambda $ & {\bf S.E.} & {\bf R2} & {\bf upper tail} $ \lambda $ & {\bf S.E.} & {\bf R2} & {\bf S.D.} \\ \hline
Low & 41.24 & 1.68 & 0.98 & 94.36 & 4.65 & 0.99 & 0.023 \\ 
Middle & 32.1 & 1.11 & 0.98 & 109.72 & 20.25 & 0.94 & 0.028 \\
high & 24.96 & 0.82 & 0.98 & 162.25 & 17.66 & 0.98 & 0.038 \\ \hline
\end{tabular}
\bigskip
\caption[Table3]{Power law exponents $ \lambda $ for the conditional probability 
distributions of the relative prices of land. The results are obtained by using OLS regression on data. S.E. denotes standard errors, and R2, the coefficient of determination.} 
\bigskip
\bigskip
\end{center}
\end{table}

To quantify the fact that the standard deviations of the relative prices vary by the price-ranges, we next calculated the standard deviation $ \sigma(s) $ of the relative prices as a function of a price. Fig. 3b shows how the standard deviation of the relative price increases linearly with the logarithm of price. Using the ordinary least squares method on the data, the standard deviation $ \sigma(s) $ is estimated as
\begin{equation}
 \sigma(s) = a + b s 
\end{equation} 
with $ a = -0.05 \pm 0.002 $ and $ b = 0.0064 \pm0.0002 $ $ R^2 = 0.996) $. The reason why the price fluctuations depend on the price level is probably as follows. The land prices in urban areas are on average higher than in rural areas, and, furthermore, trades take place more actively in urban areas than in rural ones, so that the price fluctuations are generally higher in urban areas than in rural areas. As a consequence, the price fluctuations in the high price-range are higher than the price fluctuations in the low price-range. \par 
Finally, we standardized the relative prices in order to find a form of the probability distribution of the relative prices that does not depend on the price-range. We standardized the relative price that comes into each price-range by dividing by the estimated standard deviation given a price-range. In Fig. 3c, we plot the logarithm of the probability density $ P(z | s) $ as a function of the logarithm of the standardized relative price $ z = (r - \bar{r}(s)) / \sigma(s) $, where $ \bar{r}(s) $ denotes the mean. After standardizing, the resulting empirical probability distributions appear identical for the observations drawn from different points grouped by the price-range. We found that the conditional probability distributions could be expressed by power law distributions

\begin{equation} 
P(z | s) \propto |z|^{-\lambda} 
\end{equation}
Using the ordinary least squares regression method in Fig. 3c, we obtained estimates for the asymptotic slope
\begin{equation}
\beta = \left\{ 
\begin{array}{rl} 
 0.93 \pm 0.02 & \quad\mbox{for $ z < z^* (R^2 = 0.98)$}, \\
 2.29 \pm 0.36 & \quad\mbox{for $ z \geq z^* (R^2 = 0.73)$}. 
\end{array} \right.
\end{equation}
where $ z^* = 1.76 $. To test if these results for the conditional probability distributions held for the data of the other years, we analyzed the data of each year of the 6-year period between 1995-2000. Similar quantitative behavior was found for the conditional probability distributions. 

\section{Conclusion}
We have shown that the prices and the relative prices of land are very well described by the power law distributions. Our empirical results give the conditions that any empirically accurate theories of land market have to satisfy. Since no model so far has successfully satisfied the statistical properties demonstrated in this paper, the next step is to model the behavior in land markets. In modeling land markets, we should notice that land markets are complex systems compromised of many interacting agents. In land markets, participants meet randomly and negotiated trade takes place when an agent willing to buy land meets an agent willing to sell land. The price formation in land markets is generated through this random matching process. Thus, it is likely that the methods of Statistical Physics developed to study complex systems, from which power law scaling emerges, will be useful to describe behavior in land markets. \par
A further important point to note is that land are a very important real asset, and the movement of land price strongly affect decision-making by consumers and firms [23, 24]. Therefore, the investigation of the relationship between land prices and other economic variables such as consumption and investment will lead us further into a new understanding of the dynamics of the macro-economy, which is a complex system that is composed of many interacting subsystems, each with a complex internal structure comprising many interacting agents. 

\section{Acknowledgement} 
I am indebted to Michiyo Kaizoji for her assistance in collecting the data, to Yoshi Fujiwara and Takuya Yamano for his helpful suggestions. This research was supported in part by a grant from the land institute of Japan.


\newpage
\begin{figure}
\begin{center}
  \includegraphics[height=18cm]{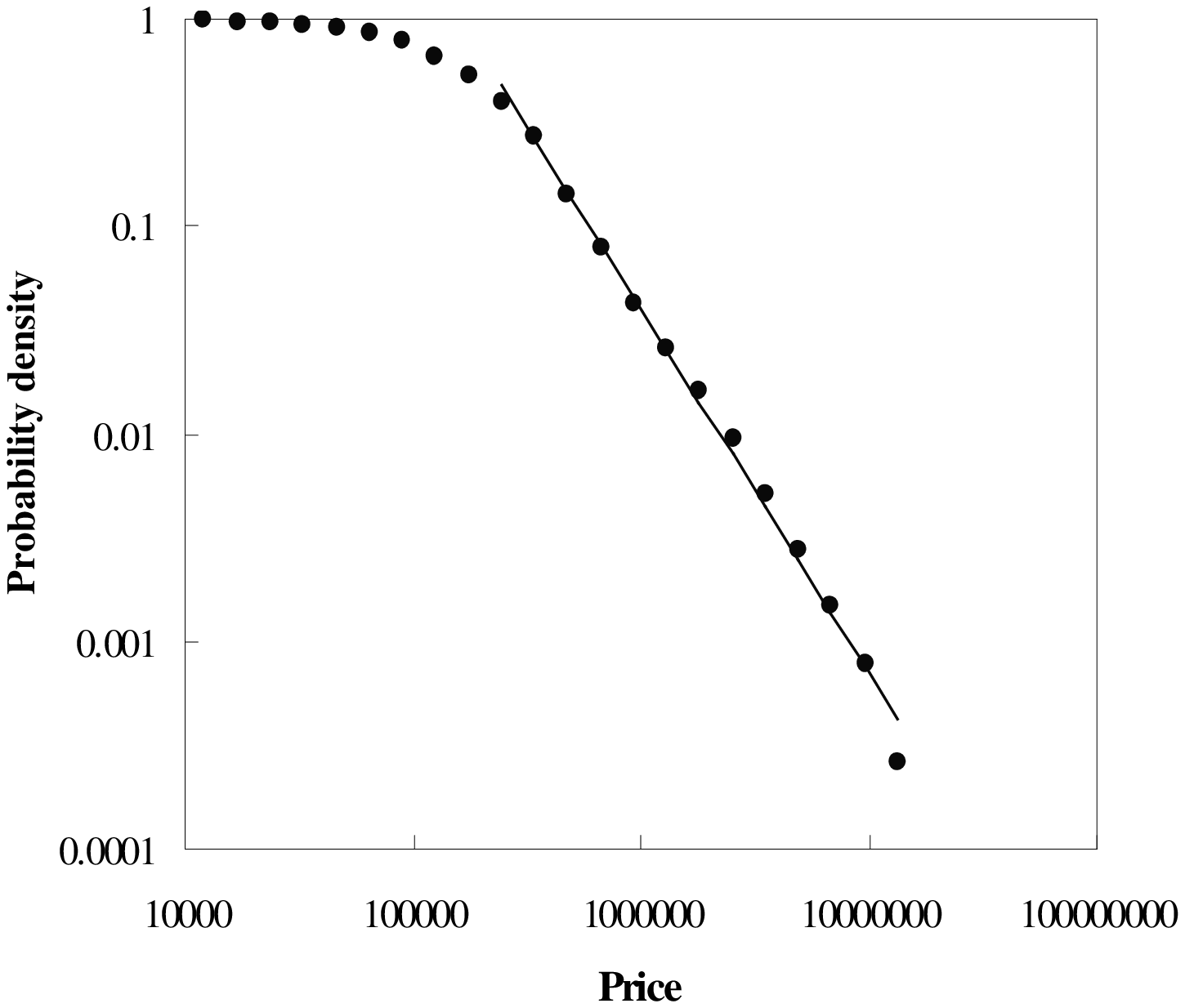}
\end{center}
\caption{Tail cumulative distribution function of land prices. Data are for 1998 from the database of the assessed value of land provided by the Ministry of Land, Infrastructure and Transport Government of Japan, the largest of which is a database of Japan's land prices in 30,600 points. The solid line is the OLS regression line in a log-log plot, with slope of $ -1.76 $ ($ SE = 0.04 $; $R^2 = 0.99 $).} 
\label{fig1}
\end{figure}

\begin{figure}
\begin{center}
  \includegraphics[width=13cm]{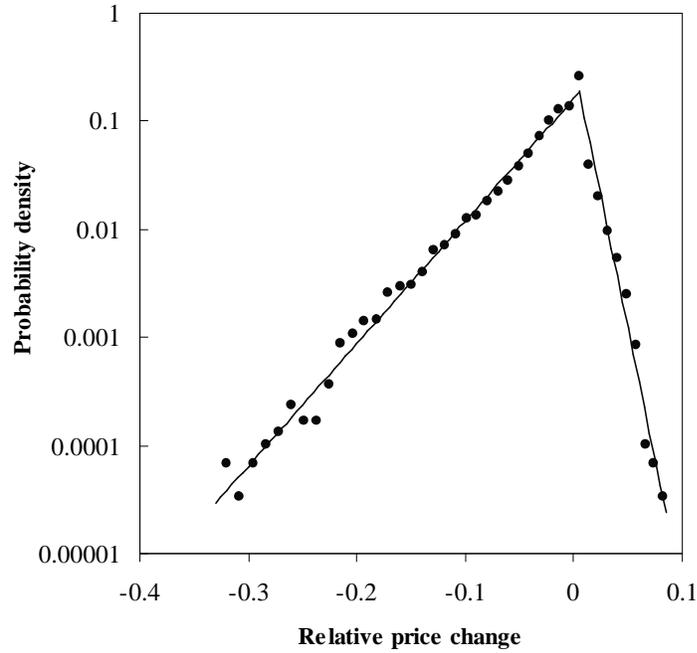}
\end{center}
\caption{Probability distribution of the relative prices on a log-log scale. Data are the ratio of prices from year 1997 to 1998 for Japan's land prices in $29,834$ points. The horizontal axis denotes the relative price change which is defined as the logarithm of the relative price, $ r = S(t+1)/S(t) $. The relative prices asymptotically follow a power law. The solid line is the regression fit in a log-log plot. The lines have the slopes with $ 26.11 \pm 0.47 $ for the lower tail and $ 112.37 \pm 6.02 $ for the upper tail.} 
\label{fig2}
\end{figure}

\begin{figure}
\begin{center}
  \includegraphics[width=13cm]{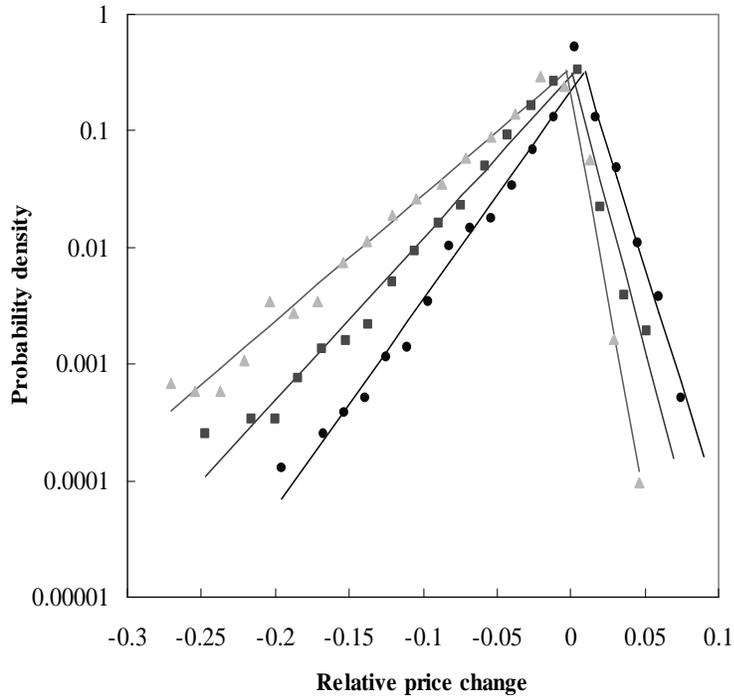}
\end{center}
\caption{The conditional probability density $ P(r | s) $ of the relative prices $ r = S(t+1) / S(t) $ from 1997 to 1998. Data contains the relative land prices at 29843 points in Japan. Different symbols refer to different price-ranges, $ s = \ln S $, $ 10.5 \leq s < 11.5 $ (circles), $ 11.5 \leq s < 12.5 $ (squares) and $ 12.5 \leq s < 13 $ (triangles). The solid lines are the OLS regression lines through the data for each of the three price-ranges in the log-log plots.}
\label{fig3}
\end{figure}

\begin{figure}
\begin{center}
  \includegraphics[width=14cm]{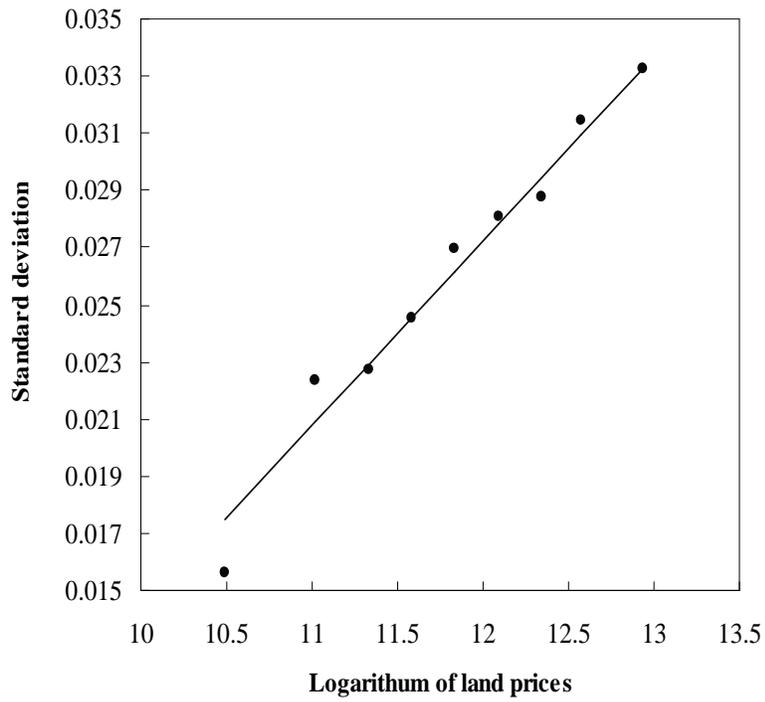}
\end{center}
\caption{ Standard deviation of the relative prices from 1997 to 1998 as a function of the logarithm of the land prices in 1997. The solid line indicates the least-square fit of equation (4) to the data.}

\label{fig4}
\end{figure}

\begin{figure}
\begin{center}
  \includegraphics[width=14cm]{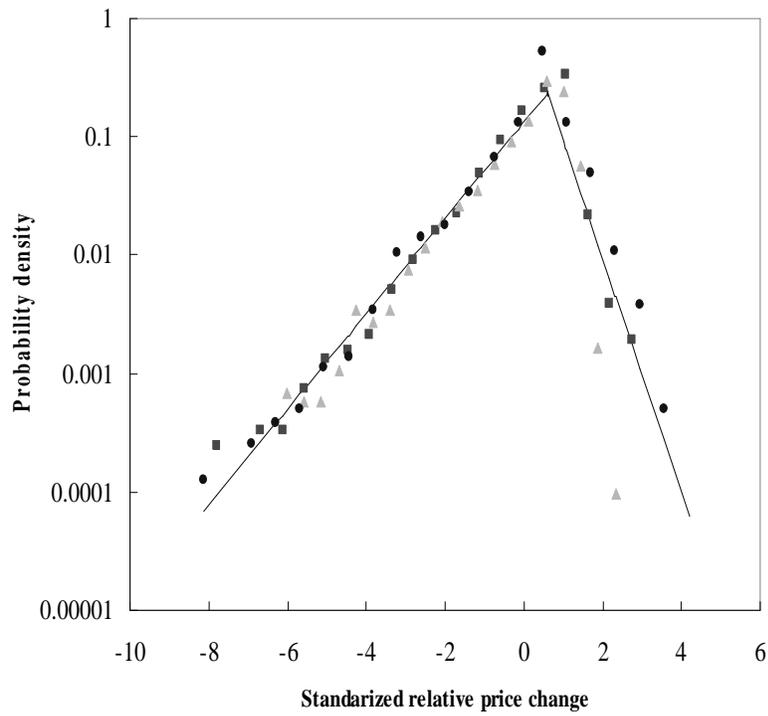}
\end{center}
\caption{Conditional probability density function $ P(z | s) $ plotted against the standardized one-year relative price $ z = (r - \bar{r}) / \sigma(s) $ for the three price-ranges defined in Fig. 3. All data collapse upon the curve (solid line) in a log-log plot. }
\label{fig5}
\end{figure}

\end{document}